\documentclass[preprintnumbers,amsmath,amssymb,floatfix,10pt,prd,onecolumn,
superscriptaddress,nofootinbib]{revtex4}
\usepackage{amsfonts}
\usepackage{mathrsfs}
\usepackage{latexsym}
\usepackage{epsfig}
\usepackage{epstopdf}
\usepackage{graphicx}
\usepackage{amssymb}
\usepackage{amsmath}
\usepackage{dcolumn}
\usepackage{bm}
\usepackage{color}
\usepackage{xcolor}
\usepackage{comment}
\usepackage[cp1251]{inputenc}
\begin{document}

\title{Investigation of Traversable Wormhole Solutions in Modified \(f(R)\) Gravity with Scalar Potential}

\author{Adnan Malik}
\email{adnanmalik_chheena@yahoo.com}\email{adnan.malik@skt.umt.edu.pk}
\affiliation{School of Mathematical Sciences, Zhejiang Normal University, \\Jinhua, Zhejiang, China.}
\affiliation{Department of Mathematics, University of Management and Technology,\\ Sialkot Campus, Lahore, Pakistan}

\author{Tayyaba Naz}
\email{tayyaba.naz@nu.edu.pk}\affiliation{National University of Computer and Emerging Sciences,\\ Lahore Campus, Pakistan.}

\author{Abdul Qadeer}
\email{abdul14114aq@gmail.com}
\affiliation{Department of Mathematics, University of Management and Technology,\\ Sialkot Campus, Lahore, Pakistan}

\author{M. Farasat Shamir}
\email{farasat.shamir@nu.edu.pk}\affiliation{National University of Computer and Emerging Sciences,\\ Lahore Campus, Pakistan.}

\author{ Zeeshan Yousaf}
\email{zeeshan.math@pu.edu.pk}\affiliation{Department of Mathematics, University of the Punjab,\\ Quaid-i-Azam Campus, Lahore-54590, Pakistan.}

\begin{abstract}
\begin{center}
\textbf{Abstract}\\
\end{center}
The objective of this manuscript is to investigate the traversable wormhole solutions in the background of the $f(R, \phi)$ theory of gravity, where $R$ is the Ricci scalar and $\phi$ is the scalar potential respectively. For this reason, we use the Karmarkar criterion for traversable static wormhole geometry to create a wormhole shape function. The suggested shape function creates wormhole geometry that links two asymptotically flat spacetime regions and meets the necessary requirements. The embedding diagram in three-dimensional Euclidean space is also discussed in order to demonstrate the wormhole configurations. For our current analysis, we choose the suitable values of free parameters for $f(R, \phi)$ gravity models to discuss the wormhole geometry. It can be observed that our proposed shape function provides the wormhole solutions with less amount of exotic matter. It can be noticed that energy conditions especially null energy conditions are violated for all considered models. The violation of energy conditions indicates the existence of exotic matter and wormhole geometry. It is concluded that the shape function acquired through the Karmarkar technique yields validated wormhole configurations with even less exotic matter correlating to the chosen $f(R, \phi)$ gravity models.
\end{abstract}
\maketitle
\date{\today}
\section{Introduction}
Exploring such types of astonishing objects as wormholes have been a fascinating topic of discussion in cosmological literature. A wormhole can be defined
as a fictitious topological feature channel that connects two distinct spacetimes in that manner; time travel will be reduced. A wormhole must have exotic fluid under the application of the General theory of relativity (GR), which tends to interrupt the null energy condition. In 1935, the well-known physicist Einstein along with Nathan Rosen suggested the concept of ``bridges" \cite{101}, which are known as the wormholes or Einstein-Rosen bridges. The goal of these bridges is to help link two disparate places in space-time, hence shortening the distance between them. It is more important to note that a wormhole is the solution of Einstein field equations in which gravity acts as ``tunnels". Wormhole is made up of two parts: a ``mouth" and a ``throat". The mouth is the entrance of the wormhole and the wormhole throat that connects the mouth. While discussing the mathematical answer for black holes, the first wormhole-like solutions were discovered. These specific answers lead to an expanded version, in which the geometric explanation is loaned to two copies of the black hole geometry. These black holes are linked by a ``throat": the Einstein-Rosen bridge or wormhole, where the throat is described as a dynamical entity dedicated to the two holes that pinch off extraordinarily and fast into a limited connection between them. A wormhole may have connected to very enormous measurements i.e. billions of light-years or more, the smallest distance being a handful of meters, connecting various places or universes in space. There are different types of wormholes mentioned over here: Schwarzschild and Kerr wormholes (Event Horizons), Morris-Thorne wormholes (Horizon-free), and Reissner-Nordstrom wormhole (Electrically charged).
Modified theories of gravities play a very important role in the investigation of wormhole geometry. Mishra, et al. \cite{2} studied the traversable wormhole geometry in $f(R)$ theory gravity, and these wormhole solutions for some assumed $f(R)$ functions are presented. The assumption of $f(R)$ is based on the fact that its behavior changed with an assumed parameter rather than the deceleration parameter. Sharif and Zahra \cite{3} explored the wormhole solutions for anisotropic and isotropic fluids as well as the barotropic equation of state with radial pressure. Benedictis and Horvat \cite{4} investigated the presence and features of wormhole throats while discussing the wormhole solutions in modified $f (R)$ theory of gravity theory\emph{}. Samanta and Godani \cite{5} calculated the range of the radius of the throat of the wormhole, where the energy conditions are satisfied. The dynamical wormhole solutions were investigated within the context of the $f(T)$ theory of gravity with anisotropic fluid, assuming a generic dynamical spherically symmetric wormhole spacetime with a specified shape function and scale factor \cite{6}. Jamil, et al. \cite{7} derived some new exact solutions of static wormholes in $f(T)$ gravity and discussed independent cases of the pressure components including isotropic and anisotropic pressure. By giving an equation of state for the matter field and applying the flaring out condition at the throat, Azizi \cite{8} was able to get the form function of the wormhole in the context of $f(R, T)$ gravity. Naz et al. \cite{9} examined the geometry of static wormholes with anisotropic matter distribution in the context of modified $f(G)$ gravity and considered the well-known Noether and conformal symmetries to investigate the geometry of the wormhole. By studying three distinct types of fluids, Malik and Nafees \cite{10} explored the spherically symmetric spacetime to debate the presence of wormhole geometry for certain actual places. Shamir and Zia \cite{11} investigated some feasible regions for the existence of traversable wormhole geometries in the background of the $f(R, G)$ theory of gravity. Samanta et al. \cite{12} have investigated a capacitative analysis of wormholes by discussing the energy conditions, equation-of-state, and anisotropy parameter are analyzed in $f(R)$ gravity, $f(R, T)$ gravity and general relativity. Sharma et al. \cite{13} investigated the solutions of traversable wormholes with normal matter in the throat within the framework of symmetric teleparallel gravity $f(Q)$, where $Q$ is the non-metricity scalar that defines the gravitational interaction. Sharif and Ikram \cite{14} formulated the explicit expressions for matter variables and evaluate wormhole solutions either specifying $f(G, T)$ model to construct the shape function or taking a specific form of the shape function to determine $f(G, T)$ model and found that null energy condition for the effective energy-momentum tensor is violated throughout the evolution in both cases while physically acceptable wormhole solutions exist only for a considered $f(G, T)$ model.

Discussion of wormhole geometry utilizing the Karmarkar condition is a very fascinating topic of interest among researchers. Karmarkar \cite{15} was the first who illustrated the compulsory condition for a static spherically symmetric spacetime to be of embedding 1. That condition is known as the Karmarkar condition which is very helpful to find the exact solutions of field equations. Fayyaz and Shamir \cite{16} constructed a wormhole shape function by using the Karmarkar condition and observed that the proposed shape function connects two asymptotically flat regions. Samanta and Godani \cite{17} investigated the traversable wormhole solutions within the background of the $f(R)$ theory of gravity with a particular viable case and discussed the energy conditions using the shape function. The same authors \cite{18} analyzed the traversable wormhole geometry and energy conditions with two different shape functions in $f(R)$ theory of gravity. Golchin and Mehdizadeh \cite{19} verified the standard energy conditions for the asymptotically spherical, flat, and hyperbolic wormhole solutions in the context of $f(R)$ modified gravity. Harko, et al. \cite{20} presented the most general conditions in the context of modified gravity, in which the matter threading the wormhole throat satisfies all of the energy conditions, and it is the higher order curvature terms, which may be interpreted as a gravitational fluid, that support these nonstandard wormhole geometries. Kuhfittig \cite{21} generalized a previous result based on the well-established embedding theorems that connect the classical theory of relativity to spacetimes of higher dimensions. The same author \cite{22} discussed the idea of embedding class one, which is applied to two diverse models, a complete solution for a charged wormhole admitting a one-parameter group of conformal motions and a new model to explain the flat rotation curves in spiral galaxies without the need for dark matter. Malik along with his collaborators \cite{23} examined the traversable wormhole models in the $ f(R) $ theories of gravity by applying the Karmarkar condition and spherically symmetric space-time. Tello-Ortiz and Contreras \cite{24} employed the class I approach to obtain wormhole solutions in the framework of general relativity in two different ways and proposed a suitable red-shift function to find its associated shape function. Gul and Sharif \cite{25} used the Noether symmetry approach to examine the viable and stable traversable wormhole solutions in the framework of the $f (R, T^2)$ theory by considering a specific model of this modified theory to obtain the exact solutions of the Noether equations. Rahaman et al. \cite{27} investigated a new wormhole solution inspired by noncommutative geometry with the additional condition of allowing conformal Killing vectors.

Modified theories of gravities may give a better explanation of the accelerating expansion of the universe. One of these modified gravities is the $f(R, \phi)$ theory of gravity, which is the mixture of the Ricci scalar $R$ and scalar potential $\phi$ \cite{28}. Malik along with his collaborators \cite{28a} discussed some dark energy cosmological models in the $f(R, \phi)$ theory of gravity, which gives a better explanation for expanding the universe. Shamir and Malik \cite{29} considered the Friedmann-Robertson-Walker spacetime for finding some exact solutions by using different values of the equation of state parameters for discussion of nature of universe. Malik et al. \cite{30} formulated the inequalities in energy constraints and use the Hubble, deceleration, jerk, and snap parameters to evaluate the feasibility of the models in the $f(R, \phi)$ theory of gravity. Malik and Shamir \cite{31} investigated the modified field equations by using anisotropic and perfect fluid distributions for the Godel-type universe in the $f(R, \phi)$ theory of gravity. Malik et al., \cite{32} studied three different Bianchi type lines elements, like Bianchi type-I, Bianchi type-III, and Kantowski Sachs space-time in the framework of $f(R, \phi)$ theory of gravity and found the exact solution of vacuum field equations by taking the valuable assumption that the expansion scalar is proportional to the shear scalar. Malik \cite{33} discussed the cylindrical symmetric solutions, especially levia-civita solutions in the background of $f(R, \phi)$ theory of gravity and examined the energy conditions in all cases. Myrzakulov et al. \cite{34} studied a class of models that can support early-time acceleration with the emerging of an effective cosmological constant at high curvature in the $f(R, \phi)$ theory of gravity. Stabile and Capozziello \cite{35} investigated the possibility of explaining theoretically the galaxy rotation curves by a gravitational potential in the total absence of dark matter. Nozari and Pourghasemi \cite{36} studied the possible crossing of the phantom divide the line in a Dvali–Gabadadze–Porrati-inspired $f(R, \phi)$ braneworld scenario and showed that there are appropriate regions of the parameter space which account for late-time acceleration and admit crossing of the phantom divide line. Panda et al. \cite{37} studied a constant-roll inflationary model in the Palatini formalism and calculated the tensor-to-scalar ratio and the spectral index using the slow-roll parameters and the results obtained are matched with the Planck 2018 data in $f(R, \phi)$ theory of gravity.

Inspired by the above past relevant work, we investigated the traversable wormhole solutions in \(f(R,\phi)\) gravity using the Karmarkar condition. The arrangement of this paper is as follows: Section III is dedicated to presenting the Class I methodology. In section III, we presented the embedding diagram for further analysis of wormhole geometry. Some basic formalism of the \(f(R,\phi)\) theory of gravity has been investigated in Section IV. In section V, we considered four compatible models of \(f(R,\phi)\) gravity and discussed the energy conditions with graphical analysis. The comparison of our current research work with past related work has been discussed in Section VI. Lastly, we have concluded our work in Section VII.

\section{FORMATION OF TRAVERSABLE WORMHOLE SOLUTION FUNCTION}
We have taken the spherically symmetric space-time as,
\begin{equation}\label{A}
 ds^2=-e^{a(r)}dt^2+e^{b(r)}dr^2+r^2 d \theta^2+r^2 \sin^2\theta d \phi^2.
\end{equation}
The non-zero Riemannian curvature components of above line element Eq. (\ref{A}) are given as,
\begin{equation}\label{B}\\
\begin{split}
R_{1313}=R_{3131},~~~~~~ R_{1212}=R_{2121},~~~~~~ R_{1221}=R_{2112},\\
R_{1331}=R_{3113},~~~~~~ R_{2332}=R_{3223},~~~~~~ R_{2323}=R_{3232},\\
R_{1414}=R_{4141},~~~~~~ R_{4242}=R_{2424},~~~~~~ R_{4114}=R_{1441},\\
R_{4224}=R_{2442},~~~~~~ R_{4334}=R_{3443},~~~~~~ R_{4343}=R_{3434}.
\end{split}
\end{equation}
Now, we define the famous known Karmarkar constraint as
\begin{equation}\label{C}
 R_{1414}=\frac{R_{1212}R_{3434}+R_{1221}R_{4334}}{R_{2323}}.
\end{equation}
By putting the values of all non-zero Riemannian curvature components in the above Karmarkar relation, we get the following equation as
\begin{equation}\label{D}
\frac{1}{2}a''e^2+\frac{1}{4}a'^2e^a-\frac{1}{4}a'b'e^b=
\frac{\frac{1}{2}(rb')(\frac{1}{2}ra'e^{a-b}\sin^2\theta)+\frac{1}{2}(rb')(\frac{1}{2}ra'e^{a-b}\sin^2\theta)}{-e^b r^2 \sin^2\theta +r^2 \sin^2\theta}.
\end{equation}

\begin{figure}[h]
    \includegraphics[width=5cm,height=4cm]{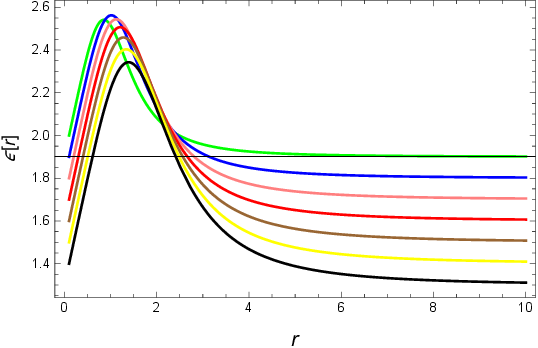}
~~~~~~\includegraphics[width=5cm,height=4cm]{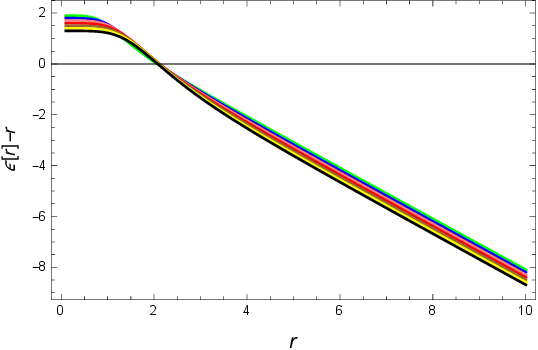}
~~~~~~\includegraphics[width=5cm,height=4cm]{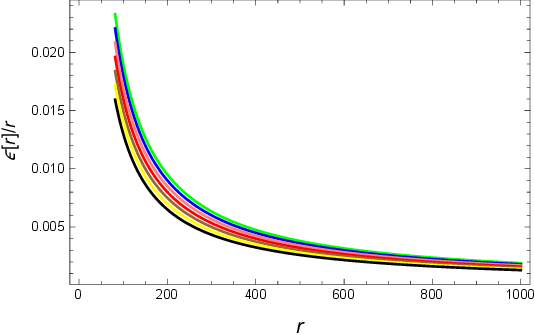}
\caption{\label{Fig.A} Graphical behavior of $\varepsilon(r)$, $\varepsilon(r)-r$ and $\varepsilon(r)/r$  from left to right.}
\end{figure}

\begin{figure}[h]
\includegraphics[width=5cm,height=4cm]{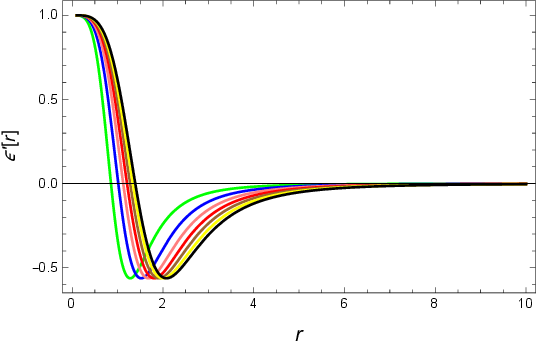}
~~~~~~~~~~~~~~~~~~\includegraphics[width=5cm,height=4cm]{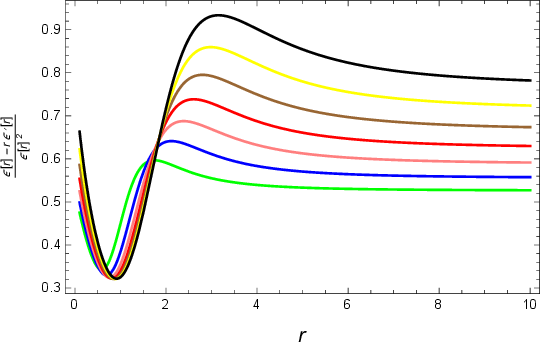}
\caption{\label{Fig.B} Graphical behavior of $\frac{\varepsilon(r)-r\varepsilon'(r)}{\varepsilon^2(r)}$ and $\varepsilon'(r)$ from left to right.}
\end{figure}

Here, $a$ and $b$ are functions of the radial coordinate $r$. By simplifying, we get the following relation as
\begin{equation*}
b'a'-a''2-a'^2= \frac{2b'a'}{-e^b+1},
\end{equation*}
where $e^b$ is not equal to 1. Now by interpreting the above equation, we get
\begin{equation}\label{E}
 e^b=B e^a a'^2+1,
\end{equation}
Here $B$ is a constant of integration. To construct wormhole shape function, we have taken the Morris-Thorne metric as
\begin{equation}\label{F}
 ds^2=-e^{a(r)}dt^2+\frac{1}{1-\frac{\varepsilon}{r}} dr^2+r^2 d \theta^2+r^2 \sin^2\theta d \phi^2.
\end{equation}
By comparing Eqs. (\ref{A}) and (\ref{F}), we obtain
\begin{equation}\label{G}
 e^b=\frac{1}{1-\frac{\varepsilon}{r}}.
\end{equation}
For further calculation, we take a redshift function $a(r)$ \cite{43} as
  \begin{equation}\label{H}
 a(r)= -\frac{2\eta}{r},
\end{equation}
where $\eta$ is a constant. Using the above Eqs. (\ref{E}) ,(\ref{G}) and (\ref{H}), we get the wormhole shape function as
  \begin{equation}\label{I}
 \varepsilon(r)=r-\frac{r^5}{r^4+4B\eta^2 e^{\frac{-2 \eta}{r}}}.
\end{equation}
By Morris and Thorne, wormhole shape function must fulfill the following important conditions:
\begin{itemize}
\item At the throat: $r=\varepsilon(r)$ at $r=r_o$ .
\item The essential condition $\frac{\varepsilon(r)-r\varepsilon'(r)}{\varepsilon'(r)} > 0$ must be satisfy at $r=r_o$.
\item $\varepsilon(r)$ must satisfy $\varepsilon'(r)<1$.
\item For the maintenance of the asymptotical flatness of
space time geometry, $\frac{\varepsilon(r)}{r}\to 0$ as $r \to \infty.$
\end{itemize}
Here, $r$ represents the radial coordinate and $r_o$ is the throat radius, and we have $r_o\leq r\leq \infty$. By evaluating Eq. (\ref{I}) at the throat condition, $\varepsilon(r_o)-r_o=0$, we get trivial solutions  at $r=r_o$. To deal with this issue, we will add an arbitrary constant C in Eq. (\ref{I}) as
\begin{equation}\label{J}
 \varepsilon(r)=r-\frac{r^5}{r^4+4B\eta^2e^{\frac{-2\eta}{r}}}+C.
\end{equation}
In the above equation, we apply the condition $r=\varepsilon(r)$ at $r=r_o$ and after some simplification, we get
\begin{equation}\label{k}
 B=\frac{r_o^4(r_o-C)}{4C\eta^2e^{\frac{-2\eta}{r}}}.
\end{equation}
By using the above value of $B$ in Eq.(\ref{J}), we get
\begin{equation}\label{L}
 \varepsilon(r)=r-\frac{r^5}{r^4+r_o^4(r_o-C)}+C.
\end{equation}
We can observe that the behavior of $\varepsilon(r)$ is positive through the graphical representation in Fig. (\ref{Fig.A}). The graphical analysis of $\varepsilon(r)-r$ is initially positive and then becomes negative with the movement of radial coordinate $r$, as shown in the second plot of Fig. (\ref{Fig.A}). Furthermore, it is clear from the third plot of Fig. (\ref{Fig.A}) that the $\frac{\varepsilon(r)}{r}\to 0$ as $r\to \infty$. Moreover, left plot of Fig. (\ref{Fig.B}) shows the satisfaction of essential condition $\frac{\varepsilon(r)-r\varepsilon'(r)}{\varepsilon'(r)} > 0$ and it can be observed that $\varepsilon'(r)<1$ is satisfied from the right plot of Fig. (\ref{Fig.B}).
\section{EMBEDDING DIAGRAM FOR WORMHOLE GEOMETRY}
In this section, we have used an embedding diagram to illustrate the behavior of wormhole geometry. To understand the working of gravity in our universe, the embedding diagram is very essential for it. Now we can set $\theta=\frac{\pi}{2}$ and $t=const$ in spherically symmetric space-time. Using this assumption in Eq. (\ref{F}), then it comes as
\begin{equation}\label{14a}
ds^2=\frac{r}{r-\varepsilon}dr^2+r^2d\phi^2.
\end{equation}
Now, we can write the above equation in three dimensional cylindrical coordinates as,
 \begin{equation}\label{15a}
ds^2=dr^2+dh^2+r^2d\phi^2, ds^2=dr^2(1+(\frac{dh}{dr})^2)+r^2d\phi^2.
\end{equation}
After comparing Eq. (\ref{14a}) and Eq. (\ref{15a}), we obtain
\begin{equation*}
\frac{r}{r-\varepsilon}=1+(\frac{dh}{dr})^2.
\end{equation*}
After simplification, we get
\begin{equation}\label{16a}
\frac{dh}{dr}=(1-\frac{\varepsilon}{r})^{\frac{-1}{2}}.
\end{equation}
An example of an embedding diagram for the bottom and top universes can be seen in Fig. (\ref{Fig.bb}) and Fig. (\ref{Fig.aa}), by utilizing $\theta=\frac{\pi}{2}$ and $t=const$. Moreover, Eq. (\ref{16a}) indicates that the embedded surface at the throat is vertical because $\frac{dh}{dr}\to \infty$. Furthermore, we have studied that how space is spherically flat apart from the throat as $\frac{dh}{dr} \to \infty$ at the time $r\to 0$. The embedded diagram is shown in Fig. (\ref{Fig.aa}) and Fig. (\ref{Fig.bb}) for the upper and lower universes, respectively, in radial coordinates with $2\pi$ rotation about the h axis.

\section{\(f(R,\phi\)) Gravity}
The action of \(f(R,\phi)\) theory of gravity is given as \cite{38}-\cite{42},
 \begin{equation}\label{2}
 S=\frac{1}{2K}\int d^4x\sqrt{-g}(f(R,\phi)+w(\phi)\phi_;\phi^{;\alpha}+L_m),
\end{equation}
where,
\begin{itemize}
  \item  $\textit{L}_m$ is lagrangian matter,
  \item  $\kappa=8\pi G$,
  \item  $g$ represents the determinant of $g_{\mu v}$,
  \item  $f(R,\phi)$ is an analytic function of Ricci scalar and scalar potential.
\end{itemize}
Now the field equation of $f(R,\phi)$ gravity can be obtained by varying the action in Eq. (\ref{2}) with respect to the metric tensor is given as,
\begin{equation}\label{3}
f_RR_{\mu v}-\frac{1}{2}[f+w(\phi)\phi_;\phi^{;\alpha}]g_{\mu v}+w(\phi)\phi_{;\mu}\phi_{;v}-f_{r;\mu v}+g_{\mu v}\square f_R =kT_{\mu v}.
\end{equation}
Here, $w$ is an arbitrary function of $\phi$, $\square=\nabla_{\mu}\nabla^v $, $\nabla_{\mu}$ denotes the covariant derivative and $f_R \equiv \frac{\partial f}{\partial R}$. The energy momentum tensor is given by,
 \begin{equation}\label{4}
T^m_{\eta \xi}=(\rho + p_t)u_\eta u_\xi -p_tg_{\eta	\xi}+(p_r-p_t)v_\eta v_\xi,
\end{equation}
where $\rho$ , $p_r$ and $p_t $ represent the energy density, radial and tangential components of pressure respectively. We have $u_\eta =e^{a/2}\delta^0_\eta$ and $v_\xi =e^{b/2}\delta^1_\xi$. Furthermore, we have considered the spherically symmetric spacetime as
 \begin{equation}\label{5}
 ds^2=-e^{a(r)}dt^2+e^{b(r)}dr^2+r^2 d \theta^2+r^2 \sin^2\theta d \theta^2.
\end{equation}
By using line element Eq. (\ref{5}), along with energy momentum tensor Eq. (\ref{4}) in Eq.(\ref{3}), we get the following field equations as
\begin{equation}\label{6}
 \rho = -f/2 + e^{-b}\bigg[-f_R''-f_R'\bigg(\frac{2}{r}+\frac{b'(r)}{2}\bigg)+\frac{1}{2}w(\phi)\phi'(r)^2+\frac{1}{4}f_R\bigg(\frac{4a'(r)}{r}+a'(r)^2-a'(r)b'(r)+2a''(r)\bigg)\bigg],
\end{equation}
\begin{equation}\label{7}
  p_r = f/2 + e^{-b}\bigg[f_R'\bigg(\frac{2}{r}+\frac{a'(r)}{2}+b'(r)\bigg)+\frac{1}{2}w(\phi)\phi'(r)^2-\frac{1}{4}f_R\bigg(a'(r)^2-\frac{4b'(r)}{r}-a'(r)b'(r)+2a''(r)\bigg)\bigg],
\end{equation}
\begin{equation}\label{8}
 p_t = f/2 + e^{-b}\bigg[-f_R''-f_R'\bigg(\frac{1}{r}+\frac{a'(r)}{2}+\frac{b'(r)}{2}\bigg)-\frac{1}{2}w(\phi)\phi'(r)^2+\frac{1}{2}f_R\bigg(\frac{-2}{r^2}+\frac{2e^{b(r)}}{r^2}-\frac{a'(r)}{2}+\frac{b'(r)}{2}\bigg)\bigg].
\end{equation}

Here, we have $w(\phi)=w_o$ $\phi^m $ and $\phi=r^\beta$, where $m , w_o$ and $\beta$ are arbitrary non-zero constants. For the further analysis and calculation, we have used the famous software MATHEMATICA.
\section{Discussion of Energy Conditions for Different Models}
In this section we have taken four different models of  \(f(R,\phi)\) gravity and investigated the nature of these constraints via graphical representation. Moreover, we explain different energy constraints and the method of predicting the existence of wormholes by using these constraints. Energy conditions are very important and helpful for explaining traversable wormhole geometry because the violation of energy conditions may predict the presence of wormholes. The energy conditions, we utilize in our study can be explained as:
\begin{eqnarray}
\nonumber NEC&:&\rho+p_r\geq 0,~~~~~~~~~~\rho+p_t\geq 0,\\
\nonumber WEC&:&\rho\geq 0,~~~~~~~~~~~~~~~~~\rho+p_r\geq 0,~~~~~~~~~~\rho+p_t\geq 0,\\
\nonumber SEC&:&\rho+p_r\geq 0,~~~~~~\rho+p_t\geq 0,~~~~~~~~~~\rho+p_r+2p_t\geq 0,\\
\nonumber DEC&:&\rho\geq 0,~~~~~~~~~~~~~~~~~\rho-|p_r|\geq 0,~~~~~~~~\rho-|p_t|\geq 0.
\end{eqnarray}
Here, NEC, WEC, SEC and DEC stand for null energy conditions, werak energy conditions, strong energy conditions and dominant energy conditions respectively. Now we will discuss the four different compatible models and investigates the nature of these energy constraints via graphical representation.
\section*{A. Model-I}
\begin{figure}[h]
 \includegraphics[width=5cm,height=4cm]{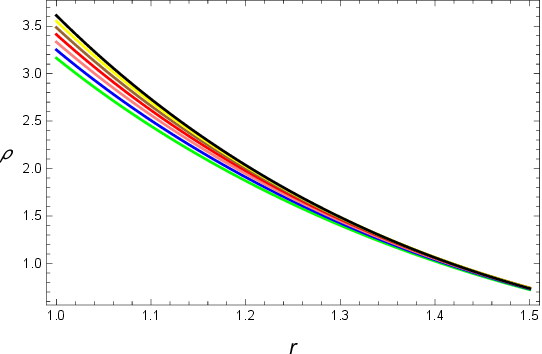}
~~~~~~\includegraphics[width=5cm,height=4cm]{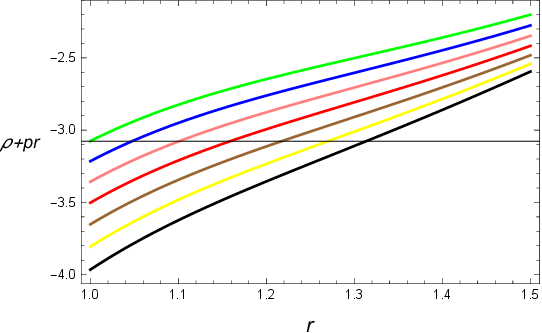}
~~~~~~\includegraphics[width=5cm,height=4cm]{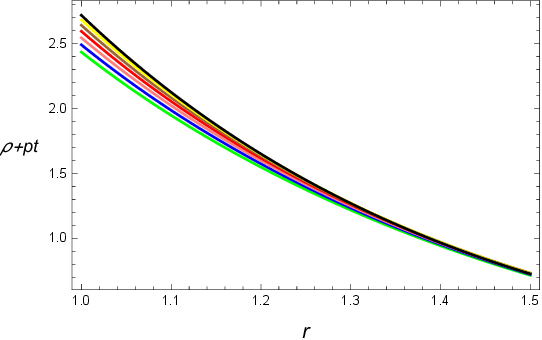}
\caption{\label{Fig.1} Shows graph of $\rho$(first plot), $(\rho+p_r)$(second plot) and $(\rho+p_t)$(third plot) from left to right.}
\end{figure}

The first model, we have considered for our analysis is,
\begin{equation}\label{9}
 f(R,\phi)=(R+\alpha R^2)*\phi(r),
\end{equation}
where $\alpha$ is free parameter and $\phi = \phi (r)$. Using this model in Eqs. (\ref{6})-(\ref{8}), we have
 \begin{equation}\label{10}
 \rho = \frac{1}{4} e^{-b}r^{-2+\beta}\bigg[-2(e^b r^2 R(1+R\alpha)+\beta(2+2\beta-(r^\beta)^{1+m}wo\beta+4R\alpha(1+\beta)))+r(1+2R\alpha)(-2\beta b'+a'(4+r a'-r b')+2 r a'')\bigg],
\end{equation}
 \begin{equation}\label{11}
 p_r = \frac{1}{4} e^{-b}r^{-2+\beta}\bigg[2(e^b r^2 R(1+R\alpha)+\beta(4+8R \alpha+(r^\beta)^{1+m}wo\beta))+r(1+2R\alpha)(-ra'^2+4b'(1+\beta)+(2\beta+ b'r)a'-2 r a'')\bigg],
\end{equation}
 \begin{equation}\label{12}
 p_t = \frac{1}{4} e^{-b}r^{-2+\beta}\bigg[e^b(2+4R\alpha+ r^2 R(1+R\alpha))-(r^\beta)^{1+m}wo\beta^2+2(1+2R\alpha)(-1+\beta^2)+r(1+2R\alpha)((-1+\beta)a'+(1+\beta)b')\bigg].
\end{equation}
According to the graphical analysis of the above Eqs. (\ref{10}-\ref{12}), it can be seen that the behavior of $\rho$ is positive and decreasing, which can be seen in Fig. (\ref{Fig.1}). Furthermore, the graphical nature of $\rho+p_t$ is positive but the nature of $\rho+p_r$ is negative, as shown in the right and middle plot of Fig. (\ref{Fig.1}). The negative trends of $\rho+p_r$ are the justification that NEC is violated. As NEC is linked with WEC, so we can also conclude that WEC is violated. Moreover, it can be observed in Fig. (\ref{Fig.2}) that the behavior of $\rho-p_r$ and  $\rho-p_t$ is positive, so DEC is satisfied. The graphical behavior of $\rho+2p_t+p_r$ is also negative which means that SEC is violated as seen in Fig. (\ref{Fig.2}). Hence, the violation of these energy constraints especially WEC and NEC indicates the presence of exotic matter, which may justify wormhole existence in the $f(R, \phi)$ gravity model.

 \section*{B. Model-II}
\begin{figure}[h]
\includegraphics[width=4cm,height=4cm]{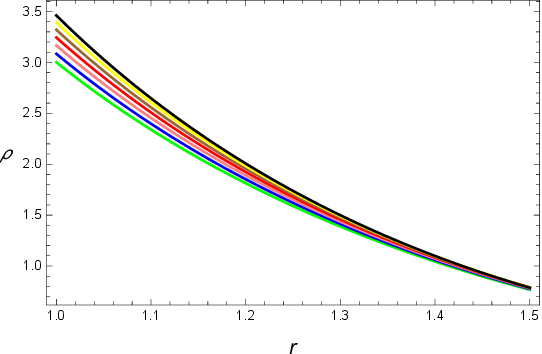}
~~~~~~\includegraphics[width=4cm,height=4cm]{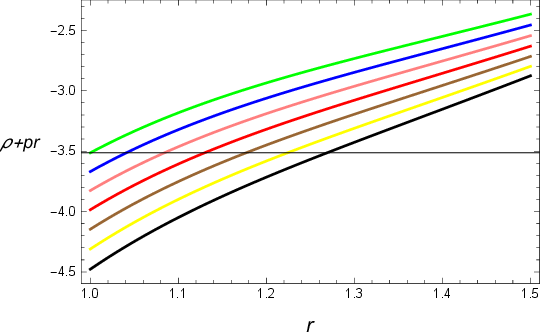}
~~~~~~\includegraphics[width=4cm,height=4cm]{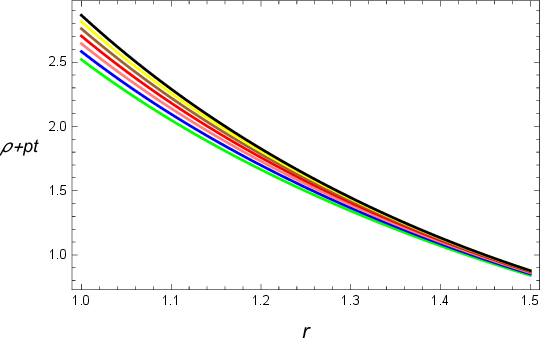}
\caption{\label{Fig.3} Shows graph of $\rho$(first plot), $(\rho+p_r)$(second plot) and $(\rho+p_t)$(third plot) from left to right.}
\end{figure}

 In this case we have taken another compatible model for the investigation of wormholes geometry. Our second model is given as
   \begin{equation}\label{13}
 f(R,\phi)=\phi(r)^M*R.
\end{equation}
 Using this model Eq. (\ref{13}) in Eq.(\ref{6})-(\ref{8}), we have:
 \begin{equation}\label{14}
 \rho = \frac{e^{-b}}{4r^2}\big[2r^{2\beta}r^{\beta M}wo\beta^2-r^{\beta M}(2e^b r^2R+4M\beta(M\beta+1)+r(2M\beta b'+a'(-4-ra'+rb')-2ra''))\big],
\end{equation}
 \begin{equation}\label{15}
 p_r = \frac{e^{-b}}{4r^2}\big[2r^{2\beta}r^{\beta M}wo\beta^2+r^{\beta M}(2e^b r^2R+8M\beta+r(-ra'^2 +4(1+M\beta) b'+a'(2M\beta+rb')-2ra''))\big],
\end{equation}
 \begin{equation}\label{16}
 p_t = \frac{e^{-b}}{2r^2}\big[-r^{2\beta}r^{\beta M}wo\beta^2+r^{\beta M}(e^b (2+r^2R)+r(-1+M\beta)a'+r(1+M\beta)(-2+2M\beta+rb'))\big].
 \end{equation}
According to the graphical analysis of the above Eqs. (\ref{14}-\ref{16}), it can be seen that the nature of $\rho$ is positive but decreasing towards the boundary as can be seen in the first plot of Fig. (\ref{Fig.3}). Furthermore, the nature of $\rho+p_t$ is positive and $\rho+p_r$ is negative, which can be seen in the right and middle plot of Fig. (\ref{Fig.3}), which represent the disturbance of NEC. We can also notice that not only NEC but WEC is also disobeyed. Besides this, it can be seen through the extreme left and middle plot of Fig. (\ref{Fig.4}) that the nature of $\rho-p_r$ and  $\rho-p_t$ is positive, so DEC is not violated in this case. The behavior of $\rho+2p_t+p_r$, which can be seen in Fig. (\ref{Fig.4}), is also negative which indicates the violation of SEC. Hence, the violation of these energy constraints specially WEC and NEC indicates the presence of wormholes in this model.

 \section*{C. Model-III}
\begin{figure}[h]
\includegraphics[width=5cm,height=4cm]{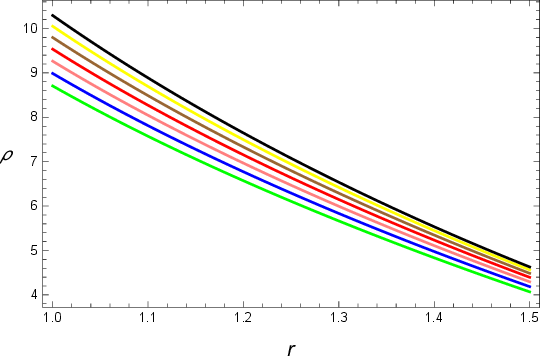}
~~~~~~\includegraphics[width=5cm,height=4cm]{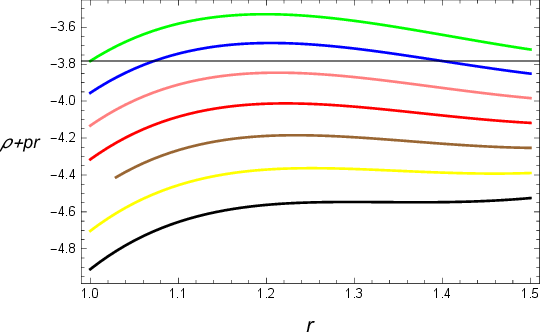}
~~~~~~\includegraphics[width=5cm,height=4cm]{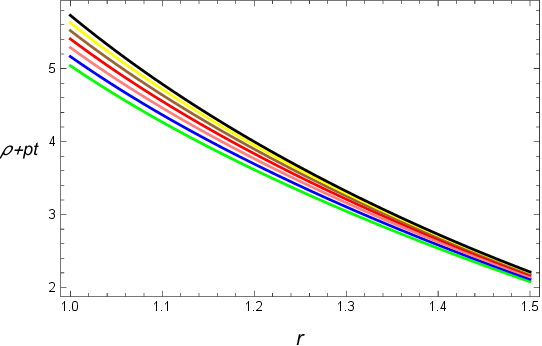}
\caption{\label{Fig.5} Shows graph of $\rho$(first plot), $(\rho+p_r)$(second plot) and $(\rho+p_t)$(third plot) from left to right.}
\end{figure}

 Here we have considered another special model which is given as,
   \begin{equation}\label{17}
 f(R,\phi)=(1+k^2 \phi(r)^2 \xi)*R,
\end{equation}
where $k, \beta$, and $\xi$ are arbitrary constants. By using the above model Eq. (\ref{17}) in Eq. (\ref{6})-(\ref{8}) as a result we obtain the following equations as:
 \begin{equation}\label{18}
 \rho = \frac{1}{4r^2}\big[-2r^2 R(1+k^2 r^{2\beta} \xi)+e^-b(r(a'(4+ra'-rb'+2ra''+r^{2\beta}(2r^{\beta m}wo\beta^2+k^2\xi(-8 \beta(1+2 \beta)+r(-4\beta b'+a'(4+ra'-rb'+2ra''))))\big],
\end{equation}
 \begin{equation}\label{19}
p_r = \frac{1}{4r}\big[2r^2 R(1+k^2 r^{2\beta} \xi)+e^-b(2r^{-1+2\beta}r^{\beta m}wo\beta^2+4k^2 r^{-1+2\beta}\beta \xi (4+ra'+2rb')+(1+k^2 r^{2\beta} \xi)(-ra'^2+4b'+ra'b'-2ra''))\big],
\end{equation}
 \begin{equation}\label{20}
 p_t = \frac{1}{2r^2}\big[e^-b(-2+2e^b+e^br^2R-ra'+rb'+r^{2\beta}(-r^{m\beta}wo\beta^2+k^2(-2+e^b(2+r^2R)+8\beta^2)\xi+k^2r\xi((-1+2\beta)a'+(1+2\beta)b')))\big].
 \end{equation}
By analyzing the graphical analysis of the above equations, we can predict the presence of wormholes for this model. With the help of the graphical analysis of the above Eqs. (\ref{14})-(\ref{16}), we can demonstrate that the nature of $\rho$ is positive but decreasing, which is given in the extreme left plot of Fig. (\ref{Fig.5}). Furthermore, for the disobeying nature of NEC, we must need at least $\rho+p_r <0 $ or  $\rho+p_t <0 $. As the behavior of $\rho+p_t$ is positive and $\rho+p_r$ is negative, can be seen in the right and middle plot of Fig. (\ref{Fig.5}) that shows the disobeying nature of NEC. We can also exhibit that not only NEC but WEC is also disobeyed because both are linked. It is shown in Fig. (\ref{Fig.6}) that the nature of $\rho-p_t$ and $\rho-p_r$ is positive, so we can observe that DEC is not disobeyed in this case. The behavior of $\rho+2p_t+p_r$, which is given in Fig.(\ref{Fig.6}), is also negative which indicates the disobeying behavior of SEC. Hence, the disturbance of these energy constraints may justify the existence of a wormhole in this model.

\section*{D. Model-IV}
\begin{figure}[h]
\includegraphics[width=5cm,height=4cm]{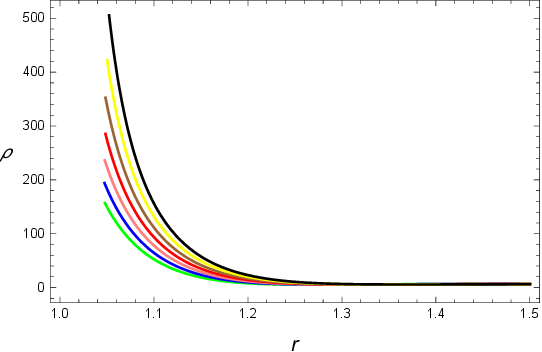}
~~~~~~\includegraphics[width=5cm,height=4cm]{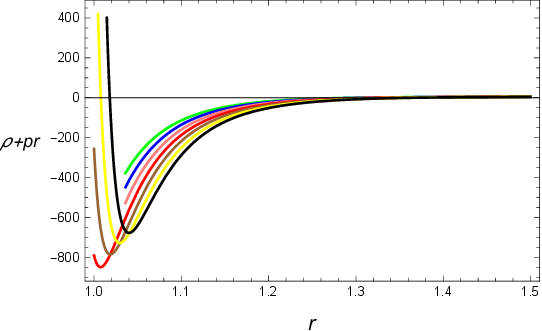}
~~~~~~\includegraphics[width=5cm,height=4cm]{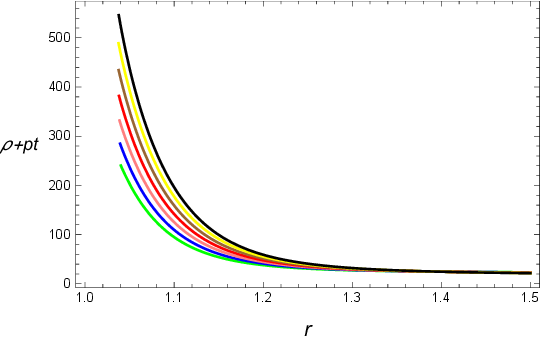}
\caption{\label{Fig.7} Shows graph of $\rho$(first plot), $(\rho+p_r)$(second plot) and $(\rho+p_t)$(third plot)  from left to right.}
\end{figure}

\begin{figure}[h]
 \includegraphics[width=5cm,height=4cm]{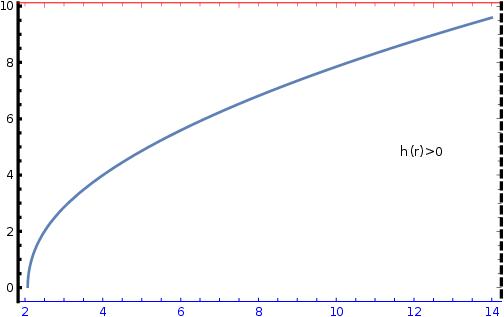}
~~~~~~~~~~~~~~~~~~~~~~~\includegraphics[width=5cm,height=4cm]{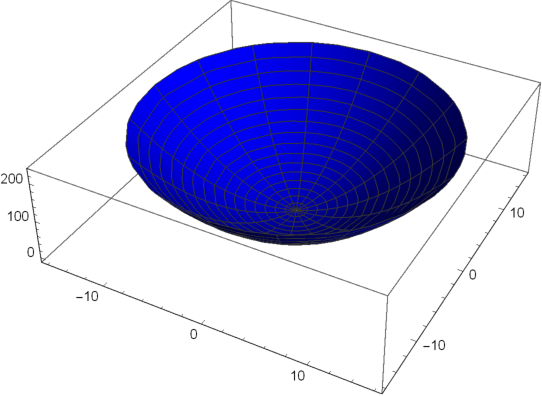}
\caption{\label{Fig.aa} Graphical representation of embedding diagram ($h(r)>0$ and it is for upper universe).}
\includegraphics[width=5cm,height=4cm]{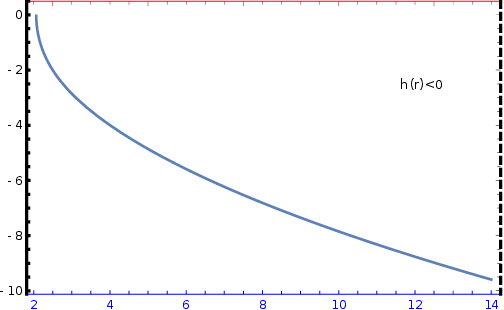}
~~~~~~~~~~~~~~~~~~~~~~~\includegraphics[width=5cm,height=4cm]{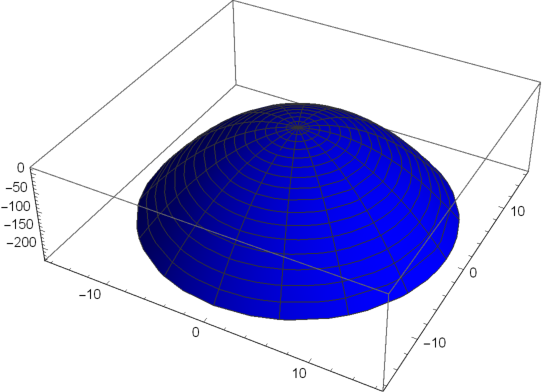}
\caption{\label{Fig.bb} Graphical representation of embedding diagram ($h(r)<0$ and it is for bottom universe).}
\end{figure}

Last model that we have considered for our investigation is given as,
   \begin{equation}\label{21}
 f(R,\phi)=\phi(r)+R^{4/3},
\end{equation}
Now by utilizing above model Eq. (\ref{21}) in Eqs. (\ref{6})-(\ref{8}), we have:
  \begin{equation}\label{22}
 \rho = \frac{1}{6r}\big[-3r(r^\beta+R^{4/3})+e^-b(3r^{-1+2\beta}r^{m\beta}wo\beta^2+2R^{2/3}(a'(4+ra'-rb')+2ra''))\big],
\end{equation}
 \begin{equation}\label{23}
p_r = \frac{1}{2}\big(r^\beta+R^{4/3})+e^-b(\frac{1}{2}r^{-2+2\beta}r^{m\beta}wo\beta^2-\frac{1}{3}R^{1/3}(a'^2-\frac{4b'}{r}-a'b'+2a'')\big),
\end{equation}
 \begin{equation}\label{24}
 p_t = \frac{1}{6r^2}\big[3r(r^\beta+R^{4/3})+e^-b(-3r^{-1+2\beta}r^{m\beta}wo\beta^2+4R^{1/3}(-2+2e^b-ra'+rb'))\big].
 \end{equation}
 With the help of the graphical analysis of the above Eqs. (\ref{22})-(\ref{24}), we can observe that the nature of $\rho$ is positive with decreasing behavior, as given in the first plot of Fig. (\ref{Fig.7}). As the nature of $\rho+p_t$ is positive and $\rho+p_r$ is negative that can be seen through the right and middle plot of Fig. (\ref{Fig.7}), which represents the disobeying behavior of NEC. We can also demonstrate that not only NEC but WEC is also showing disobeying behavior because whenever NEC disobeyed WEC must be violated. Besides this, it can be seen from the left and middle plot of Fig. (\ref{Fig.8}) that the graphical analysis of $\rho-p_r$ is positive and  $\rho-p_t$ shows positive but also negative behavior near the boundary, so we can conclude that DEC is disobeyed. The behavior of $\rho+2p_t+p_r$, which is given in Fig.(\ref{Fig.8}), is also negative which indicates the violation of SEC. Hence, the disobeying nature of these energy constraints, particularly NEC and WEC may justify the existence of a wormhole in this model.

\section{Comparison}
Shamir and fayyaz \cite{112} constructed a wormhole shape function by employing the Karmarkar condition for static traversable wormhole geometry and generated the wormhole geometry that connected two asymptotically flat regions of spacetime and satisfied the required conditions. In our work, we have discussed the embedding diagram in three-dimensional Euclidean space to present the wormhole configurations and investigate the traversable wormhole geometry in $f(R, \phi)$ gravity. Moreover, the authors \cite{112} considered three different models to discuss the wormhole geometry in $f(R)$ gravity but we have considered four different models in $f(R,\phi)$ gravity to investigate the behavior of traversable wormholes, which make our work more generalized and comprehensive than the previous investigation. Godani and Samanta \cite{aa} discussed the wormhole solutions in $f(R)$ theory of gravity by taking well defined $f(R)$ gravity model and particular shape functions. However, in this work we first construct the shape function by using the Karmarkar condition and then discuss the wormhole solutions for more than one model, which is the interesting feature of our work. Sharif and Fatima \cite{a1} evaluated the traversable wormhole solutions through Karmarkar condition in $f(R, T)$ theory of gravity. They considered two different $f(R, T)=f(R)+f(T)$ gravity models, where $f(R)$ was taken as exponential gravity model and Starobinsky model but $f(T)=\lambda T$ was fixed for the sake of simplicity. In our present work, we investigate the traversable wormhole solutions for $f(R, \phi)$ gravity models by taking the product of Ricci scalar $R$ and scalar potential $\phi$, which make our work more interesting with the previous investigations. Banerjee et al. \cite{a1} presented a class of solutions for static and spherically symmetric wormholes geometries by considering specific choices for the $f (Q)$ form and for constant redshift with different shape functions. We have done analysis for wormhole solutions by constructing the wormhole geometry, and investigate the energy conditions for all considered cases. Furthermore, we perform a more detailed analysis by taking different values of $C$ appeared in the shape function i.e., $C=1.9$ ~(color Green), $ C=1.8$ ~(Blue color), $ C= 1.7$~ (color Pink), $ C= 1.6$~ (color Red), $ C= 1.5$~ (color Brown), $ C= 1.4$ ~(color Yellow), $ C= 1.3$ ~(color Black). This is also a very crucial point of our research work, which make our work different from the previous one. Mishra, et al. \cite{a3} discussed the traversable wormhole geometry by taking three shape functions and constant redshift but our analysis is different because we choose a more general redshift function depending upon the radial coordinate $r$.
\begin{figure}[h]
\includegraphics[width=5cm,height=4cm]{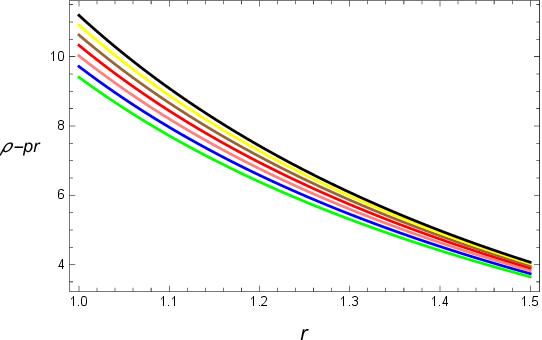}
~~~~~~\includegraphics[width=5cm,height=4cm]{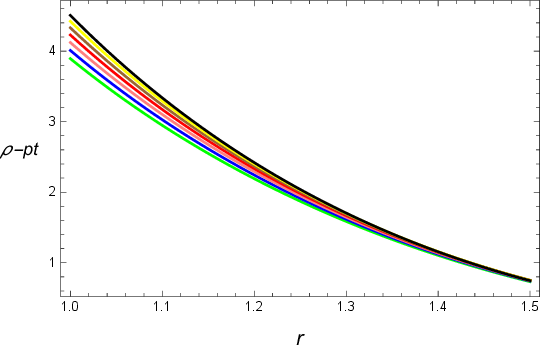}
~~~~~~\includegraphics[width=5cm,height=4cm]{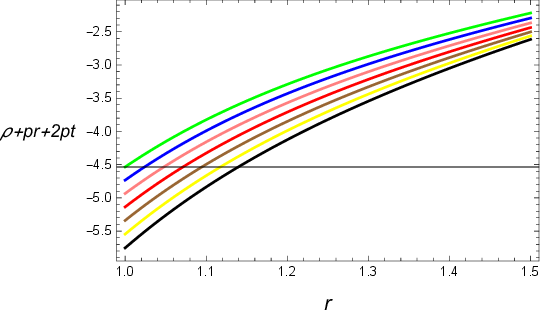}
\caption{\label{Fig.2} Shows graph of $(\rho-p_r)$(first plot), $(\rho-p_t)$(second plot) and $(\rho+2p_t+p_r)$(third plot) from left to right.}
\end{figure}
\begin{figure}[h]
\includegraphics[width=5cm,height=4cm]{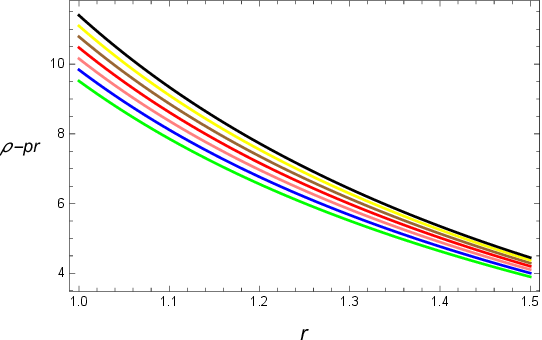}
~~~~~~\includegraphics[width=4cm,height=4cm]{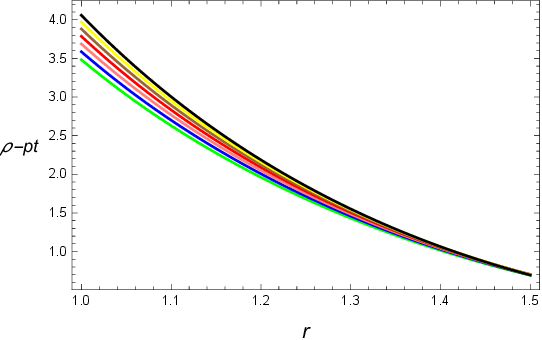}
~~~~~~\includegraphics[width=4cm,height=4cm]{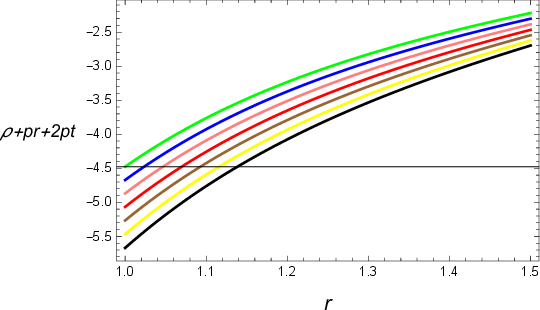}
\caption{\label{Fig.4} Shows graph of $(\rho-p_r)$(first plot), $(\rho-p_t)$(second plot) and $(\rho+2p_t+p_r)$(third plot) from left to right.}
\end{figure}
\begin{figure}[h]
\includegraphics[width=5cm,height=4cm]{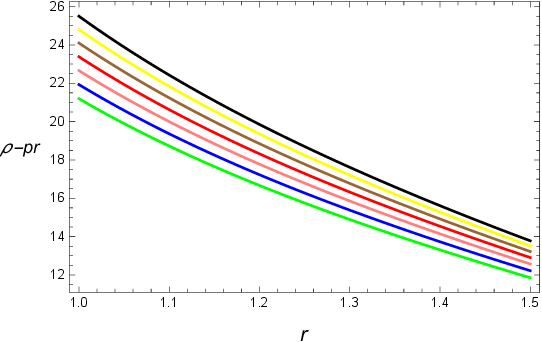}
~~~~~~\includegraphics[width=5cm,height=4cm]{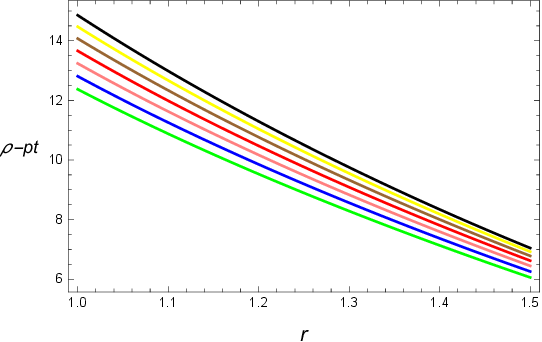}
~~~~~~\includegraphics[width=5cm,height=4cm]{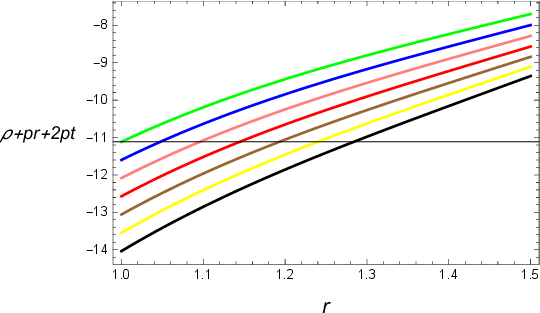}
\caption{\label{Fig.6} Shows graph of $(\rho-p_r)$(first plot), $(\rho-p_t)$(second plot) and  $(\rho+p_r+2p_t)$(third plot) from left to right.}
\end{figure}
\begin{figure}[h]
\includegraphics[width=5cm,height=4cm]{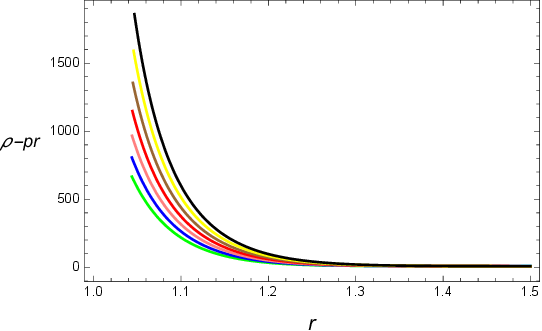}
~~~~~~\includegraphics[width=5cm,height=4cm]{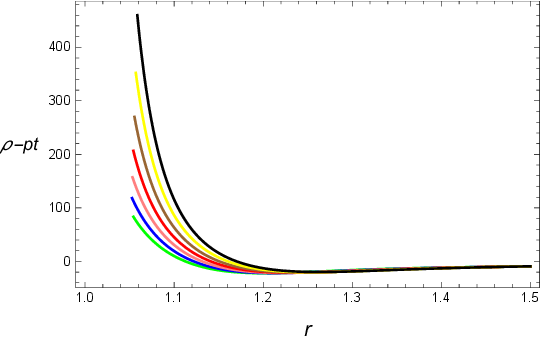}
~~~~~~\includegraphics[width=5cm,height=4cm]{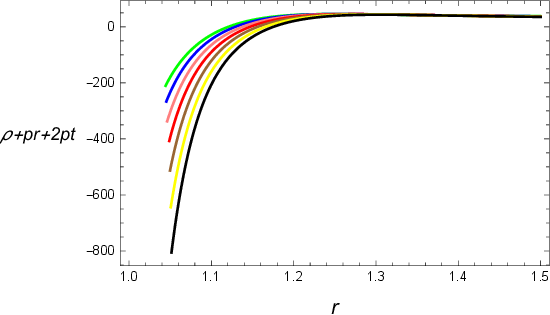}
\caption{\label{Fig.8} Shows graph of  $(\rho-p_r)$(first plot), $(\rho-p_t)$(second plot) and $(\rho+2p_t+p_r)$(third plot) from left to right.}
\end{figure}
\section{Concluded remarks}
The purpose of this research is to study the wormhole geometry by using the Karmarkar condition in the $f(R,\phi)$ theory of gravity. For our current analysis, we have used special four models of $f(R,\phi)$ theory of gravity and investigated the validity of different types of energy constraints.
It is worthwhile to mention here that modified gravity theories with a scalar field may accommodate a wide range of dark energy and modified gravity models \cite{a4}. The addition of scalar field helps to modify the cosmological and galactic dynamics leaving the solar system unaffected.
As far as wormhole study is concerned, the addition of scalar field supports any asymptotically well-behaved traversable wormhole. In our work, scalar field helps in violating NEC which is necessary for the existence of exotic matter and wormhole geometry. The stable behavior of wormhole is also supported by the addition of scalar field, though we have not discussed this in present work. To the best of our knowledge, it is the first attempt to explore wormhole solutions in $f(R,\phi)$ gravity with Karmarker condition. The detailed outcomes of our research work are as follows:

The graphical representation of energy density is positive but decreasing towards the boundary as can be seen in the first plot of Fig. (\ref{Fig.1}). The graphical nature of $\rho+p_t$ is positive but $\rho+p_r$ is negative, as seen in Fig. (\ref{Fig.1}). It can be noticed that NEC and WEC are violated due to the negative behavior of the $\rho+p_r$ component. Moreover, the graphical representation of $\rho-p_r$ and $\rho-p_t$ is positive, which means that DEC is satisfied as seen in Fig. (\ref{Fig.2}). The graphical trends of $\rho+2p_t+p_r$ are negative, which is the justification that SEC is violated as seen in Fig. (\ref{Fig.2}). Hence, the violation of these energy constraints specially WEC and NEC indicates the presence of wormholes in this model.

According to the graphical analysis for Model II, it can be seen that the nature of energy density is positive and becomes decreasing, when we move on the radial coordinate as plotted in Fig. (\ref{Fig.3}). It can also be noticed that NEC is violated due to the negative nature of $\rho+p_r$ as seen in Fig. (\ref{Fig.3}). It can also be observed that WEC is also violated because NEC is violated. For our second model, DEC is satisfied because $\rho-p_r>0$ and $\rho-p_t>0$, but SEC is violated due to $\rho+2p_t+p_r<0$, as seen in Fig. (\ref{Fig.4}). The significant consequences of energy condition violation, especially violation of the NEC, may indicate the presence of exotic matter, which may justify wormhole existence in the $f(R,\phi)$
gravity model.

In our third model, we observe that the graphical trends of energy density are maximum near the origin and positive as seen in Fig. (\ref{Fig.5}). Furthermore, it can be seen that $p_t+\rho$ is positive but $p_r+\rho$ is negative, which is the justification that NEC is violated. We can also conclude that not only NEC but WEC is also violated because both are connected. It is also noticed that the nature of $\rho-p_r$ and  $\rho-p_t$ is positive, so DEC is again satisfied in our third model. It can be seen from Fig. (\ref{Fig.6}) that the nature of $\rho+2p_t+p_r$ is negative, which is a violation of SEC.

In our fourth and last model, the behavior of energy density is also positive with decreasing nature as shown in Fig. (\ref{Fig.7}). Furthermore, the behavior of $p_t+\rho$ is positive and $p_r+\rho$ is negative, which indicates the violation of NEC and WEC. Besides this, it can be seen from Fig. (\ref{Fig.8}) that the graphical analysis of $\rho-p_r$ is positive and  $\rho-p_t$ shows positive but also negative behavior near the boundary, so we can conclude that DEC is disobeyed. The behavior of $\rho+2p_t+p_r$, which is given in Fig.(\ref{Fig.8}), is also negative which indicates the violation of SEC. Hence, the disobeying nature of these energy constraints, particularly NEC and WEC may justify the existence of a wormhole in this model. From above discussion, we can conclude that in above all four models these energy constraints are violated. Since these energy constraints are very important in the study of wormholes. The significant consequences of energy condition violation, especially violation of the NEC, may indicate the presence of exotic matter, which may justify wormhole existence in the $f(R,\phi)$ gravity models.

Finally, it is stated that the shape function acquired through the Karmarkar technique yields validated wormhole configurations with even less exotic matter correlating to the proper choice of $f(R, \phi)$ gravity models and acceptable free parameter values. As a nutshell, the present models fulfill all of the criterion for the presence of a wormhole. In particular, it is concluded that the gravitational field emerging from higher order terms of modified gravity with scalar field favor the existence of the wormhole geometries.

\section*{Data Availability Statement}
\hskip\parindent
\small
No data was used for the research in this article. It is pure mathematics.
\section*{Conflict of Interest}
\hskip\parindent
\small
The authors declare that they have no conflict of interest.

\section*{Contributions}
\hskip\parindent
\small
 We declare that all the authors have same contributions to this paper.
\section*{ACKNOWLEDGMENTS}
\hskip\parindent
\small
Adnan Malik acknowledges the Grant No. YS304023912 to support his Postdoctoral Fellowship at Zhejiang Normal University, China.

  \end{document}